\documentstyle[11pt,aaspp4]{article}

\begin{document}

\title{
	Three-micron imaging of the Hubble Deep Field\altaffilmark{1}
}
\author{
	David~W.~Hogg,\altaffilmark{2,3}
	Gerry~Neugebauer,\altaffilmark{4}
	Judith~G.~Cohen,\altaffilmark{4}
	Mark~Dickinson,\altaffilmark{5}
	S.~G.~Djorgovski,\altaffilmark{4}
	Keith~Matthews,\altaffilmark{4}
	B.~T.~Soifer\altaffilmark{4}
}
\altaffiltext{1}{
Based on observations made at the W. M. Keck Observatory, which is
operated jointly by the California Institute of Technology and the
University of California; and with the NASA/ESA Hubble Space
Telescope, which is operated by AURA under NASA contract NAS 5-26555.
}
\altaffiltext{2}{
Institute for Advanced Study, Olden Lane, Princeton NJ 08540, USA;
{\tt hogg@ias.edu}
}
\altaffiltext{3}{
Hubble Fellow
}
\altaffiltext{4}{
Palomar Observatory, California Institute of Technology,
mail code 105-24, Pasadena CA 91125, USA
}
\altaffiltext{5}{
Space Telescope Science Institute,
3700 San Martin Drive, Baltimore MD 21218, USA
}

\begin{abstract}
Images of the Hubble Deep Field at $3.2~{\rm \mu m}$, taken with the
10-m Keck Telescope, are presented.  The images cover a total area of
$\sim 2.5~{\rm arcmin^2}$.  To a 5-sigma limit of $[3.2]_{\rm
total}\approx 17.5~{\rm mag}$ (Vega-relative), 11 sources are
detected, 9 of which are extragalactic.  The integrated galaxy number
counts are therefore $\sim 1.3\times 10^4~{\rm deg^{-2}}$ to this
depth.  The galaxies detected at $3.2~{\rm \mu m}$ have a median
redshift of $z= 0.56$.  All $3.2~{\rm \mu m}$ sources have $1.6~{\rm
\mu m}$, $1.1~{\rm \mu m}$ and visual counterparts, all of fairly
regular morphologies; several also have $6.7~{\rm \mu m}$, $15~{\rm
\mu m}$, $850~{\rm \mu m}$, $8.5~{\rm GHz}$, or $1.4~{\rm GHz}$
counterparts.  No sources are found which are either anomalously red
or anomalously blue in their $H-[3.2]$ color, and there are
significant detections of most of the known near-infrared HDF sources
for which detection in these $3.2~{\rm \mu m}$ data seemed likely.
\end{abstract}


\section{Introduction}

Until now, it has been impractical to make wide-field ground-based
surveys at infrared wavelengths longer than $2.4~{\rm \mu m}$.
Although the atmosphere is transparent in wavelength windows in the
near infrared out to $10~{\rm \mu m}$, sky brightness increases
rapidly with increasing wavelength.  Detector area is limited by the
need to read out quickly before saturation.  The Near Infrared Camera
(NIRC; Matthews \& Soifer 1994) on the 10-m W.~M.~Keck Telescope,
however, is equipped with a fast readout mode which allows it to read
out its entire array quickly enough to avoid saturation in a
short-wavelength $L$-band filter, centered on $3.2~{\rm \mu m}$.

The $3.2~{\rm \mu m}$ bandpass is complementary to the standard
near-infrared $J$, $H$, and $K$ ($1.2$, $1.6$ and $2.2~{\rm \mu m}$)
bandpasses in a number of respects.  It admits light at the rest-frame
wavelengths of these bandpasses at redshifts of 1.7, 1.0, and 0.5, so
observations of intermediate and high-redshift galaxies can be
compared directly to local analogs observed in the standard
bandpasses.  The $3.2~{\rm \mu m}$ bandpass can be used to extend the
near-infrared spectral energy distributions (SEDs) to longer
wavelengths.  While stellar populations are made up of a superposition
of basically thermal radiators, nuclear activity tends to be redder
than thermal (in the near infrared, bluer in the visual) and thus
ought to become more prominent in redder bandpasses.  Nuclear activity
may also heat nuclear dust to produce additional near-infrared
emission (eg, Peletier et al 1999).  In units of $f_{\nu}$, the SEDs
of normal galaxies peak at $\approx 1.6~{\rm \mu m}$, primarily due to
a minimum in H$^-$ opacity in stellar atmospheres.  At $z = 1$, this
peak is redshifted to $3.2~{\rm \mu m}$.  A ``non-evolving''
$L^{\ast}$ elliptical galaxy at redshift $z=1$ ought to appear as a
$\sim 17.5$~mag source at $3.2~{\rm \mu m}$.  At $z=1$, old stellar
populations are $1.5$ to $3$ mag brighter than young stellar
populations with the same apparent $V$ magnitude (Coleman et al 1980,
Aaronson 1977).

This paper presents the results of deep imaging of the Hubble Deep
Field (HDF; Williams et al 1996) in the $3.2~{\rm \mu m}$ bandpass
with NIRC.  Although neither the sky coverage nor the number of
detected sources is large, to our knowledge this is the first deep
near-infrared survey at wavelengths beyond the $K$ band.  It extends
the SEDs of HDF sources over a decade in wavelength from the F300W
bandpass at $0.3~{\rm \mu m}$.  The choice of the HDF for deep imaging
in this new $3.2~{\rm \mu m}$ bandpass was clear: The HDF is currently
the most well-studied field on the sky, deeply imaged by almost every
astronomical technique.

In what follows, all magnitudes are Vega-relative.

\section{Observations and reduction}

The $L_{\rm short}$ filter on NIRC was chosen for this study.  The
filter passes light in the wavelength range $2.5<\lambda<3.5~{\rm \mu
m}$.  When combined with typical atmospheric absorption conditions on
Mauna Kea, the bandpass is concentrated in the wavelength range
$2.9<\lambda<3.5~{\rm \mu m}$, with a small ``blue leak'' ($\sim
10~{\rm percent}$ of the total transmission function) in a narrow peak
at $2.5~{\rm \mu m}$.  The central wavelength is $\sim 3.2~{\rm \mu
m}$ for the filter plus atmosphere transmission function; for this
reason, magnitudes with this instrument and filter setup will be
denoted $[3.2]$.  The $L_{\rm short}$ filter is currently the
longest-wavelength bandpass in which NIRC can read out its full
$256\times 256~{\rm pixel}$ ($38\times38~{\rm arcsec^2}$) field area
without saturation from sky brightness alone.  A source with
$[3.2]=0~{\rm mag}$ in the $L_{\rm short}$ bandpass has a flux density
of 330~Jy.  This approximate zeropoint is an interpolation between
measured $K$ and $L$-band zeropoints, and is uncertain by no more than
10~percent.

\begin{figure}
\plotone{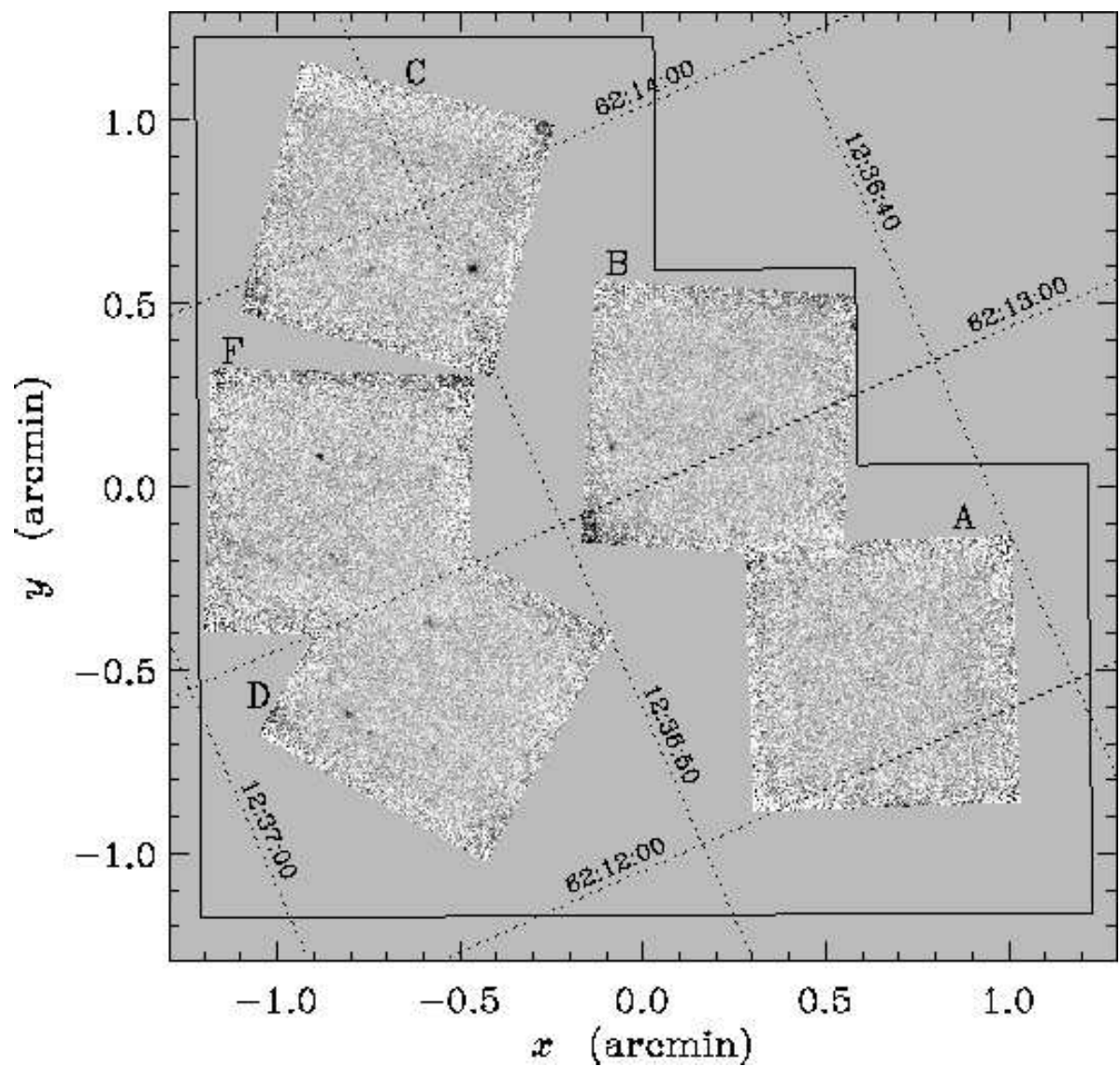}
\caption[The final $3.2~{\rm \mu m}$ mosaic]{The final $3.2~{\rm \mu
m}$ mosaic.  The subfields are labeled with letters. The outline of
the HST--WFPC2 imaged field is shown with a solid line.  Lines of
constant RA and Dec are shown and labeled.}
\label{fig:mosaic}
\end{figure}

The selection of the HDF for deep study with HST is described
elsewhere (Williams et al 1996).  Within the HST-imaged portion of the
HDF, a set of subfields were chosen for deep study with NIRC.  These
subfields were not chosen with regard to any particular ``interesting
sources.''  Rather, one subfield was chosen to lie roughly in the
center of each WFPC2 CCD chip image, subject to guide star constraints
(these are the primary subfields A, B, C, and D), and additional
subfields were chosen to fill in the gaps between these four primary
subfields.  In the end, telescope time and weather restricted our
study to five subfields, the primary four and subfield F, which lies
between subfields C and D.  The well-observed solid angle is roughly
$0.5~{\rm arcmin^2}$ per subfield (given dithering) or a total of
$2.5~{\rm arcmin^2}$.  The layout of the subfields is shown in
Figure~\ref{fig:mosaic}, relative to the HST-imaged field.

Air mass of the observations ranged from 1.40 to 1.75. The conditions
were clear and apparently photometric for all nights, although there
was some variation in the sky brightness.  Each individual exposure
consisted of 1000 (or, occasionally, 2000) 0.025~s or 0.029~s
integrations coadded into a single image in hardware.  A visual
imaging offset guider pointing at a star maintained the tracking of
the telescope.  The observations were taken in half-hour or hour-long
sets with random dithers over a $10\times 10~{\rm arcsec^2}$ range.
The photometric sensitivity was determined by measuring faint standard
stars (Elias et al 1982, Persson et al 1998) roughly once per hour
during each night.  During several of the nights, the measured
sensitivity (corrected for airmass) varied in an apparently linear way
with time.  The subfield most affected by this was subfield A, which
shows no significant sources (see below), so the variation in this
subfield does not affect the results.  The scatter of the individual
standard-star measurements around the linear sensitivity trends is
$\pm 0.05~{\rm mag}$, but for the other four subfields, the trends
themselves range between $\pm <0.01~{\rm mag}$ for subfield D and $\pm
0.12~{\rm mag}$ for subfield F.  The uncertainty in the photometric
calibration is thus no better than $0.05~{\rm mag}$, but may be worse
if the linear trends represent a serious photometric problem.  The
observing dates, integration times, and sensitivities of the final,
stacked images of the subfields are given in
Table~\ref{tab:subfields}.

The data were processed by subtracting an average of the two adjacent
images (before and after) from each image to eliminate artifacts
introduced by varying sky brightness and field rotation resulting from
the alt--azimuth configuration of the Keck Telescope.  Because the
individual images show no significant sources above the sky noise,
registration of the individual images was accomplished with telescope
coordinates as computed from the positions of the star in the guider
camera.  The individual registered images were averaged, after dark
subtraction and flat-fielding, to produce the final images shown in
Fig~\ref{fig:mosaic}.  The pixel-to-pixel rms values in the central
parts of the images correspond to $\approx 20.2$~mag in $1~{\rm
arcsec^2}$; the specific values for each subfield are given in
Table~\ref{tab:subfields}.

Because there are so few sources visible per subfield (none for
subfield A), determination of the seeing is difficult.  Seeing values
given in Table~\ref{tab:subfields} are determined by comparison of the
sources in the final, stacked images to their visual counterparts in
the WFPC2 image of the HDF taken in the F814W bandpass (Williams et al
1996); they are necessarily quite uncertain.  Since in several fields
(especially F) the seeing in the final, stacked image is apparently
worse than that for the standard stars, some of the seeing must arise
from poor registration of the frames coadded in the stacks.

For visual images, the ``version 2'' reductions of the HST/WFPC2
imaging of the HDF were used.  Details of the data acquisition,
reduction, and sensitivities are given elsewhere (Williams et al
1996).  For brevity, the F300W, F450W, F606W, and F814W bandpasses
will be referred-to here as $U$, $B$, $V$, and $I$, although F300W and
F606W are substantially different from conventional $U$ and $V$.

For other near-infrared data, the ``version 1.04'' reductions of the
imaging of the HDF taken with the NICMOS instrument on HST were used.
Details of the data acquisition, reduction, and sensitivities are
given elsewhere (Dickinson et al 2000).  For brevity, the F110W and
F160W bandpasses will be referred-to here as $J$ and $H$, although
F110W is substantially bluer than standard $J$.  The photometric
sensitivities of the $J$ and $H$ data are uncertain at the $\sim
5$~percent level.

\section{Source selection and photometry}

Sources were identified in the $3.2~{\rm \mu m}$ image by requiring a
positive deviation of greater significance than $5.0\,\sigma$ inside a
focal-plane aperture of $1.0~{\rm arcsec}$ diameter in the final,
stacked images of the subfields.  The $1\,\sigma$ noise levels were
estimated by measuring fluxes in randomly placed apertures, so they
correctly take into account any pixel-to-pixel correlations.  At
significances less than $10\,\sigma$, it was established that the
sources had visual and near-infrared counterparts in the HST images.
No $5.0\,\sigma$ deviations failed to show counterparts, except at the
very edges of the subfields, where the noise and background flatness
are degraded.  A total of 11 sources were selected.  The total area
subject to source detection is roughly $2.5~{\rm arcmin^2}$.

Total magnitudes $[3.2]_{\rm total}$ for the sources selected at
$3.2~{\rm \mu m}$ were estimated by measuring aperture magnitudes
$[3.2]_{\rm ap}$ in focal-plane apertures of $2.0~{\rm arcsec}$
diameter and then correcting for the small size of the aperture.  The
$2.0~{\rm arcsec}$ aperture was chosen to be small enough to reduce
the background at $3.2~{\rm \mu m}$ but large enough to reduce
variations in photometric sensitivity measured from the standard stars
to below $0.1~{\rm mag}$.  The aperture corrections were determined
under the assumption that there are only small color gradients in
$(H-[3.2])$ color, by comparing $2.0$ and $6.0~{\rm arcsec}$ diameter
aperture magnitudes in the very high signal-to-noise $H$-band image,
smoothed to the seeing of the $3.2~{\rm \mu m}$ image.  (The one
exception is the star at $12\,36\,56.368$ +$62\,12\,41.14$, for which
a $5.0~{\rm arcsec}$ diameter aperture was used in order to avoid
light from the neighboring bright galaxy.)  Aperture and total
magnitudes for the sources are given in Table~\ref{tab:catalog}.  The
uncertainties given in Table~\ref{tab:catalog} are based only on the
photon statistics and the uncertainties in determining the sky level,
and do not include uncertainties in the photometric sensitivity.  No
estimate is made for the uncertainties in the aperture corrections.
These may be considerable, if the seeing estimates are in error, if
there are significant color gradients, or if there is significant
light outside of the $6.0~{\rm arcsec}$ diameter aperture.

Colors were measured in $2.0~{\rm arcsec}$ diameter focal-plane
apertures in the visual WFPC2 and near-infrared NICMOS images,
smoothed to the seeing of the $3.2~{\rm \mu m}$ images.  The aperture
size was chosen so the colors were measured in the regions of the
sources which are well-measured at $3.2~{\rm \mu m}$.  The colors are
given in Table~\ref{tab:colors}.  The photon and read-noise
contributions to the uncertainties in the color measurements are at or
below the $0.01~{\rm mag}$ level for the all the visual and
near-infrared colors except for $(H-[3.2])$ where the uncertainties
are dominated by the uncertainty in $[3.2]_{\rm ap}$ given in
Table~\ref{tab:catalog} and $(U-B)$ where the uncertainties are at the
$0.01$ to $0.2$~mag level, depending on $U$-band flux.  Although the
NICMOS $J$ and $H$ measurements have a very high signal-to-noise
ratio, the photometric sensitivities are uncertain at the $\sim
10$~percent level (Dickinson et al 2000).  Figure~\ref{fig:colors}
shows $(V-[3.2])$ and $(H-[3.2])$ colors for the galaxies as a
function of redshift.  Figure~\ref{fig:thumbnail} shows the visual and
near-infrared counterparts of all the sources.

Absolute RA and Dec positions were assigned to the sources under the
assumption that the absolute astrometry of the visual HST images of
the HDF (Williams et al 1996) is correct.  These absolute positions
are given in Table~\ref{tab:catalog}.

\section{Comparison with other HDF catalogs}

The catalog selected at $3.2~{\rm \mu m}$ can be compared with other
catalogs made in the HDF at other wavelengths or with other
techniques.  The redshifts and flux densities at other wavelengths
discussed in this Section are presented in Tables~\ref{tab:colors} and
\ref{tab:counterparts}.

Redshifts were found by comparing the coordinates of the sources with
the lists of redshifts obtained in the Hubble Deep Field with the Keck
Telescope (Cohen et al 1996; Lowenthal et al 1997; Cohen et al 2000).
Two of the sources are Galactic stars, consistent with their stellar
appearance on the HST images.

Flux densities in the $K_s$ band ($2.15~{\rm \mu m}$) were found by
matching the absolute positions of the $3.2~{\rm \mu m}$ sources with
the positions of the sources from the $K_s$-selected catalog of Hogg
et al (2000).  Flux densities at $6.7$ and $15~{\rm \mu m}$ were found
by matching with the catalogs made from imaging data from the Infrared
Space Observatory (Aussel et al 1999; based on data from Serjeant et
al 1997).  Flux densities at $850~{\rm \mu m}$ were taken from the
catalog made from the SCUBA instrument at JCMT (Hughes et al 1998),
although the sources were identified in part through comparing
positions derived from $1.4~{\rm GHz}$ observations (Richards 1999a).
Flux densities at $8.5$ and $1.4~{\rm GHz}$ were found by matching
positions in catalogs made with VLA data at these frequencies
(Richards et al 1998; Richards 1999b).

The detection fraction at $3.2~{\rm \mu m}$ of known near-infrared
sources was tested by identifying sources in the NICMOS/HDF catalogs
(Dickinson et al 2000) satisfying
\begin{equation}
[3.2]_{\rm predicted} = F160W - 1.85\,(F110W-F160W) + 0.25 < 18
\end{equation}
which corresponds to an approximate linear extrapolation (in log-log
space) of the $J$ and $H$ magnitudes to $[3.2]$.  (This equation is
roughly appropriate for predicting the fluxes of galaxies and nuclear
emission, since both have SEDs which resemble power-laws at $3.2~{\rm
\mu m}$ (observed) at the redshifts of interest.)  All nine sources
inside the region of the HDF well-observed at $3.2~{\rm \mu m}$ with
$[3.2]_{\rm predicted}<17.0$ have counterparts in the catalog selected
at $3.2~{\rm \mu m}$, and of the the six sources in the well-observed
region with $17.0<[3.2]_{\rm predicted}<18.0$, two have counterparts
in the catalog.

\section{Results}

Images of the Hubble Deep Field at $3.2~{\rm \mu m}$, taken with the
10-m Keck Telescope were presented, representing the first
near-infrared survey of the distant Universe at wavelengths beyond the
$K$ band.  The images cover a total area of $\sim 2.5~{\rm arcmin^2}$.

To a 5-sigma limit of $[3.2]_{\rm total}\approx 17.5~{\rm mag}$
(Vega-relative), 11 sources are detected, 9 of which are
extragalactic.  The integrated galaxy number counts are therefore
$\sim 1.3\times 10^4~{\rm deg^{-2}}$ to this depth.  The sources
detected at $3.2~{\rm \mu m}$ have a median redshift of $z=0.56$.

\begin{figure}
\plotone{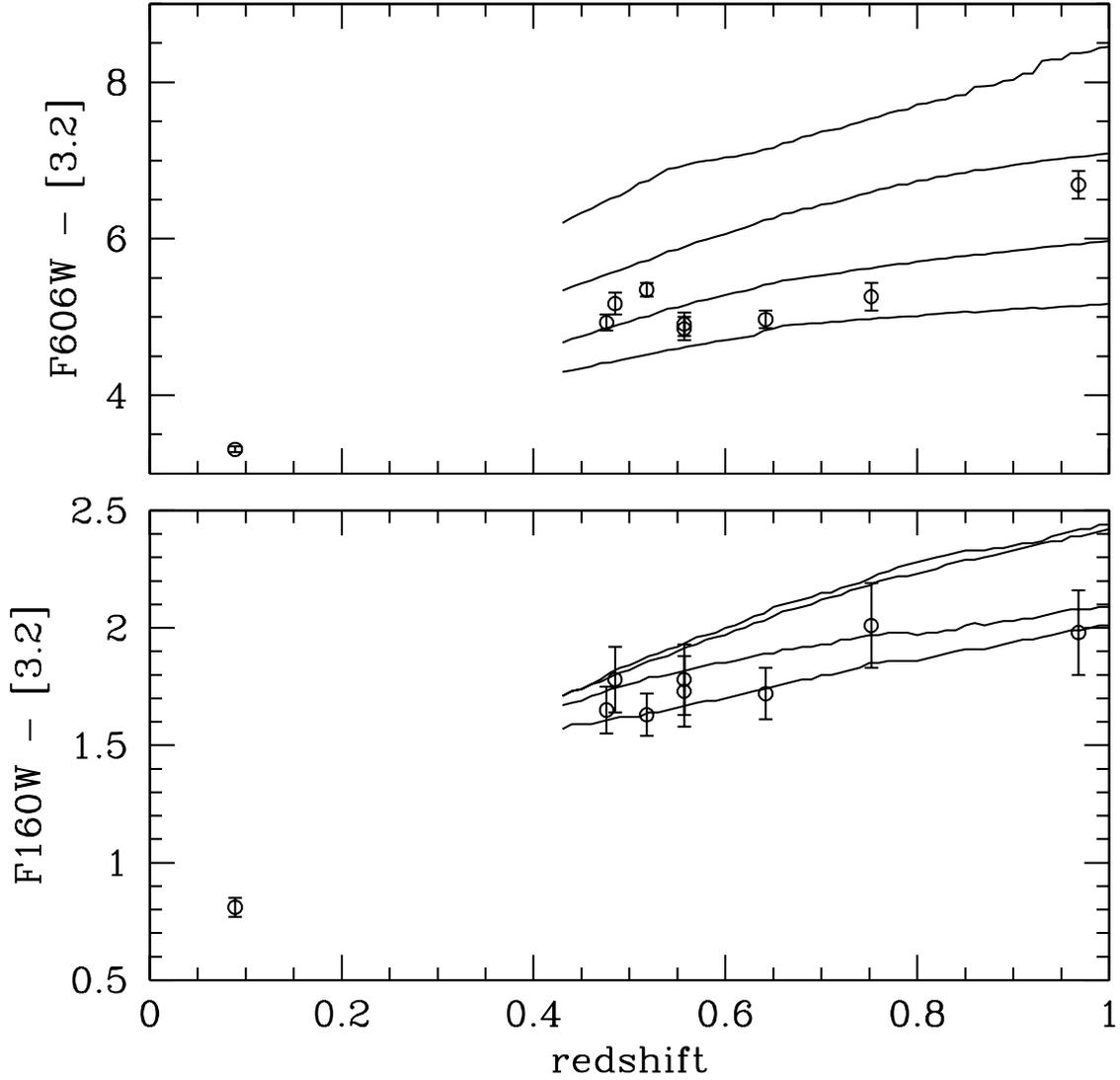}
\caption[The colors]{The $(V-[3.2])$ and $(H-[3.2])$ colors as a
function of redshift for the sources selected at $3.2~{\rm \mu m}$.
Also shown are no-evolution predictions for galaxies of types (from
reddest to bluest or top to bottom) E/S0, Sbc, Scd, and Im, based on
the bright-galaxy samples of Coleman et al (1980) and Aaronson (1977).
The predicted color-redshift tracks begin at redshift $z=0.43$ because
the bright-galaxy data do not extend longward of the $K$ band.}
\label{fig:colors}
\end{figure}

All $3.2~{\rm \mu m}$ sources have $1.6~{\rm \mu m}$, $1.1~{\rm \mu
m}$ and visual counterparts.  No sources are found which are either
anomalously red or anomalously blue in their $(V-[3.2])$ or
$(H-[3.2])$ colors.  In Figure~\ref{fig:colors}, the source colors are
compared with no-evolution predictions, based on observations of local
galaxies (Coleman et al 1980, Aaronson 1977) and simple models of the
filter transmission curves.  Although the sources selected at
$3.2~{\rm \mu m}$ are among the reddest visually-selected galaxies,
they are blue relative to no-evolution predictions.  The blue colors
are likely a consequence of higher star formation rates at
intermediate redshift than in the present-day Universe, as has been
inferred, among other ways, from metallicity in Lyman-alpha clouds
(Pei \& Fall 1995), ultraviolet luminosity density (Lilly et al 1996;
Connolly et al 1997; Madau et al 1998) and emission line strengths
(Hammer et al 1997; Heyl et al 1997; Small et al 1997; Hogg et al
1998).  The lack of very red sources does not constrain the abundance
of ``extremely red objects'' (EROs) because, at the magnitudes of
interest, the angular number density of EROs is on the order of
$10^{-2}~{\rm arcmin^{-2}}$ (Hu \& Ridgeway 1994; Thompson et al
1999).

\begin{figure}
\plotone{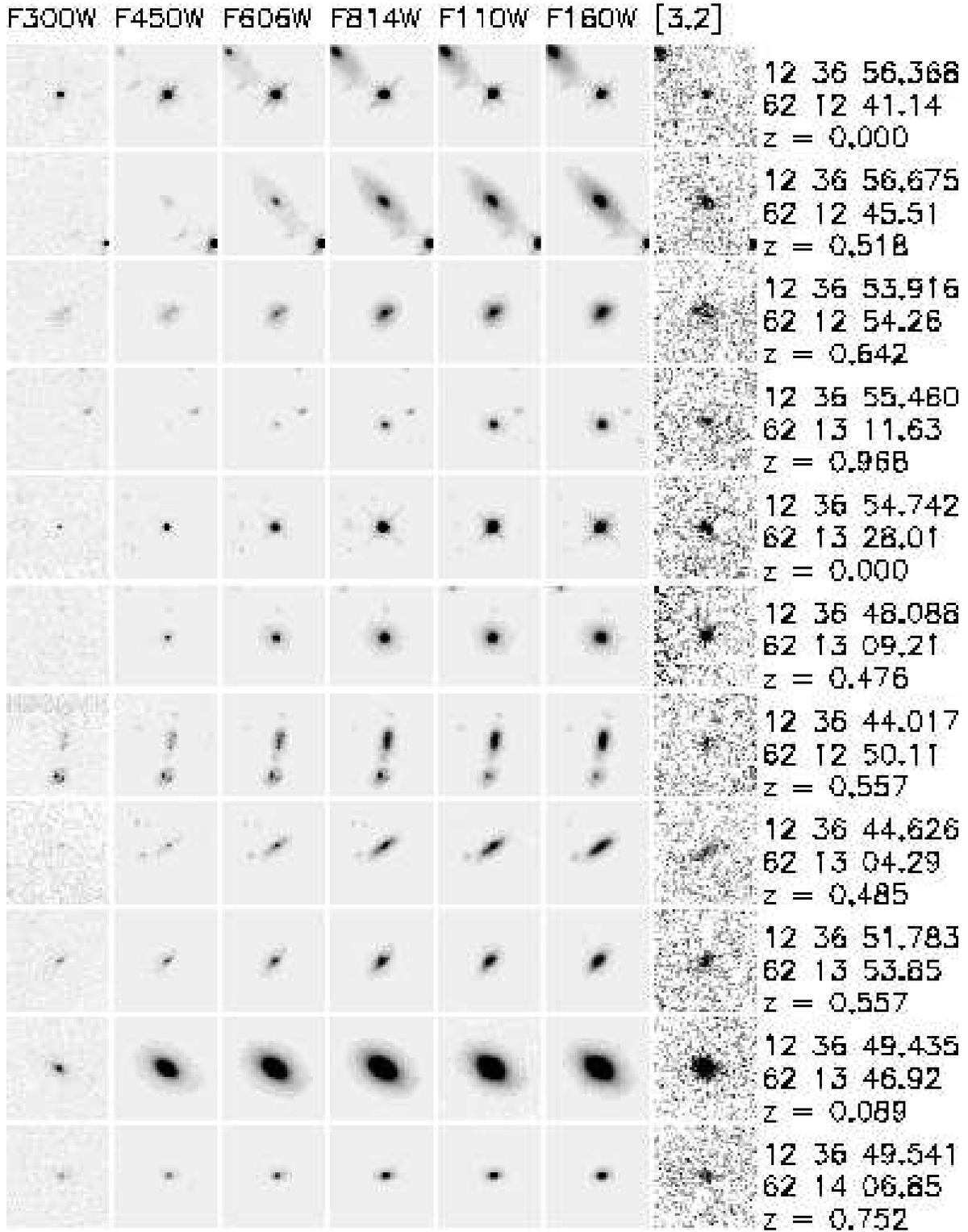}
\caption[The $2\times 2~{\rm arcsec}$ region around each source]{The
$2\times 2~{\rm arcsec}$ region around each source selected at
$3.2~{\rm \mu m}$.  The images are stretched identically in terms of
flux density per logarithmic interval in frequency $(\nu\,f_{\nu})$.}
\label{fig:thumbnail}
\end{figure}

The morphologies of the $3.2~{\rm \mu m}$ sources in the visual and
near-infrared bandpasses are shown in Figure~\ref{fig:thumbnail}.  The
sample is small, but they appear fairly regular, with some showing
nearby neighbors or evidence for mild interaction.

Comparisons with surveys of the HDF at other wavelengths show that all
of the $3.2~{\rm \mu m}$ sources have $2.15~{\rm \mu m}$ counterparts,
one has a $6.7~{\rm \mu m}$ counterpart, several have $15~{\rm \mu m}$
counterparts, one has a likely $850~{\rm \mu m}$ counterpart, and two
have $8.5~{\rm GHz}$ counterparts.  Furthermore, there are $3.2~{\rm
\mu m}$ detections of the majority of sources which, on the basis of
their $1.1~{\rm \mu m}$ and $1.6~{\rm \mu m}$ photometry, ought to
have been detected.

In short, the $3.2~{\rm \mu m}$ imaging of the HDF shows no great
surprises.  It does demonstrate that it is possible, with significant
telescope time, to perform studies of normal galaxies at high
redshifts in the $L$ band of atmospheric transmission.

\acknowledgements It is a pleasure to thank the P3 Team, in particular
Al Conrad and Bob Goodrich, for the work on the NIRC instrument which
made this study possible, John Gathright for technical support, the
HDF/NICMOS GO team for providing near-infrared data in advance of
publication, and Eric Richards for providing radio data in advance of
publication.  This research is based on observations made at the
W. M. Keck Observatory, which is operated jointly by the California
Institute of Technology and the University of California; and with the
NASA/ESA Hubble Space Telescope, which is operated by AURA under NASA
contract NAS 5-26555.  Financial support was provided by the NSF, and
by grants HF-01093.01-97A (Hubble Fellowship) and GO-07817.01-96A
(NICMOS data) from STScI, which is operated by AURA under NASA
contract NAS~5-26555.  This research made use of the NASA ADS Abstract
Service, the NOAO/IRAF data reduction software, and the SM plotting
software.


\clearpage
\begin{deluxetable}{llccc}
\tablewidth{0pt}
\tablecaption{
        Properties of the final, stacked images of the subfields
	\label{tab:subfields}
}
\tablehead{
   \colhead{name}
 & \colhead{date}
 & \colhead{exposure}
 & \colhead{seeing FWHM\tablenotemark{a}}
 & \colhead{$[3.2]$ at $1\,\sigma$\tablenotemark{b}}
\\
   \colhead{}
 & \colhead{(UT)}
 & \colhead{(s)}
 & \colhead{(arcsec)}
 & \colhead{(mag)}
}
\startdata
NIRC-HDF-A & 1999 April 28   & 13440 & 0.5\tablenotemark{c} & 20.1 \nl
NIRC-HDF-B & 1998 February 6 & 11100 & 0.5 & 20.4 \nl
NIRC-HDF-C & 1998 February 7 &  9600 & 0.4 & 20.2 \nl
NIRC-HDF-D & 1999 April 30   &  9480 & 0.4 & 20.3 \nl
NIRC-HDF-F & 1999 April 27   & 11100 & 0.6 & 20.0 \nl
\enddata
\tablenotetext{a}{Because there are so few significant sources per
subfield, the seeing measurements are very uncertain; see text.}
\tablenotetext{b}{The $1\,\sigma$ magnitude is that corresponding
to a $1\,\sigma$ variation in the sky in a focal-plane aperture of
$1~{\rm arcsec^2}$ in solid angle, in the central (i.e., full
exposure-time) part of the image.}
\tablenotetext{c}{The seeing FWHM for subfield A is taken from the
observations of the standard stars, because there are no significant
sources detected in that subfield.}
\end{deluxetable}

\begin{deluxetable}{ccccccccccc}
\tablewidth{0pt}
\tablecaption{
        The catalog selected at $3.2~{\rm \mu m}$
	\label{tab:catalog}
}
\tablehead{
   \colhead{RA\tablenotemark{a}}
 & \colhead{Dec\tablenotemark{a}}
 & \colhead{sub-}
 & \colhead{$[3.2]_{\rm ap}$\tablenotemark{b}}
 & \colhead{$[3.2]_{\rm total}$\tablenotemark{c}}
\\
   \multicolumn{2}{c}{(J2000)}
 & \colhead{field}
 & \colhead{(mag)}
 & \colhead{(mag)}
}
\startdata
$12\,36\,44.017$ & +$62\,12\,50.11$ & B & 
$17.25\pm 0.15$ & $16.70$ \nl
$12\,36\,44.626$ & +$62\,13\,04.29$ & B & 
$17.46\pm 0.14$ & $16.90$ \nl
$12\,36\,48.088$ & +$62\,13\,09.21$ & B & 
$16.78\pm 0.10$ & $16.51$ \nl
$12\,36\,49.435$ & +$62\,13\,46.92$ & C & 
$15.75\pm 0.04$ & $15.44$ \nl
$12\,36\,49.541$ & +$62\,14\,06.85$ & C & 
$17.57\pm 0.18$ & $17.43$ \nl
$12\,36\,51.783$ & +$62\,13\,53.85$ & C & 
$17.25\pm 0.15$ & $16.99$ \nl
$12\,36\,53.916$ & +$62\,12\,54.26$ & D & 
$17.11\pm 0.11$ & $16.84$ \nl
$12\,36\,54.742$ & +$62\,13\,28.01$ & F & 
$16.80\pm 0.11$ & $16.71$ \nl
$12\,36\,55.460$ & +$62\,13\,11.63$ & F & 
$17.40\pm 0.18$ & $17.07$ \nl
$12\,36\,56.368$ & +$62\,12\,41.14$ & D & 
$17.80\pm 0.21$ & $17.72$ \nl
$12\,36\,56.675$ & +$62\,12\,45.51$ & D & 
$16.85\pm 0.09$ & $16.19$ \nl
\enddata
\tablenotetext{a}{Absolute positions were found with information from
the HST/WFPC2 images (Williams et al 1996).}
\tablenotetext{b}{Aperture magnitudes $[3.2]_{\rm ap}$ are measured in
a $2.0~{\rm arcsec}$ diameter focal-plane aperture.  The quoted
uncertainty in magnitude is simply that due to uncertainty in the
photon statistics and sky level in the focal-plane aperture.  The
quoted uncertainty does not take into account uncertainty in overall
calibration.}
\tablenotetext{c}{Total magnitudes $[3.2]_{\rm total}$ have been
corrected to an effective $6.0~{\rm arcsec}$ diameter aperture with
information from the $H$ image; see text for details.}
\end{deluxetable}

\begin{deluxetable}{ccccccccccc}
\scriptsize
\tablewidth{0pt}
\tablecaption{
        Redshifts and colors of the $3.2~{\rm \mu m}$ sources
	\label{tab:colors}
}
\tablehead{
   \colhead{RA}
 & \colhead{Dec}
 & \colhead{$[3.2]_{\rm total}$}
 & \colhead{$z$\tablenotemark{a}}
 & \colhead{$U-B$\tablenotemark{b}}
 & \colhead{$B-V$\tablenotemark{b}}
 & \colhead{$V-I$\tablenotemark{b}}
 & \colhead{$I-J$\tablenotemark{b}}
 & \colhead{$J-H$\tablenotemark{b}}
 & \colhead{$H-[3.2]$\tablenotemark{b,c}}
\\
   \multicolumn{2}{c}{(J2000)}
 & \colhead{(mag)}
 & \colhead{}
 & \colhead{(mag)}
 & \colhead{(mag)}
 & \colhead{(mag)}
 & \colhead{(mag)}
 & \colhead{(mag)}
 & \colhead{(mag)}
}
\startdata
$12\,36\,44.017$ & +$62\,12\,50.11$ & $16.70$ & 0.557 &
$-0.17$ & $ 1.26$ & $ 1.15$ & $ 0.66$ & $ 1.26$ & $ 1.78$ \nl
$12\,36\,44.626$ & +$62\,13\,04.29$ & $16.90$ & 0.485 &
$ 0.28$ & $ 1.58$ & $ 1.29$ & $ 0.81$ & $ 1.29$ & $ 1.78$ \nl
$12\,36\,48.088$ & +$62\,13\,09.21$ & $16.51$ & 0.476 &
$ 1.22$ & $ 1.93$ & $ 1.38$ & $ 0.67$ & $ 1.23$ & $ 1.65$ \nl
$12\,36\,49.435$ & +$62\,13\,46.92$ & $15.44$ & 0.089 &
$ 1.46$ & $ 1.19$ & $ 0.95$ & $ 0.56$ & $ 0.99$ & $ 0.81$ \nl
$12\,36\,49.541$ & +$62\,14\,06.85$ & $17.43$ & 0.752 &
$-0.82$ & $ 0.96$ & $ 1.34$ & $ 0.67$ & $ 1.24$ & $ 2.01$ \nl
$12\,36\,51.783$ & +$62\,13\,53.85$ & $16.99$ & 0.557 &
$-0.34$ & $ 1.21$ & $ 1.21$ & $ 0.69$ & $ 1.28$ & $ 1.73$ \nl
$12\,36\,53.916$ & +$62\,12\,54.26$ & $16.84$ & 0.642 &
$-0.55$ & $ 1.12$ & $ 1.31$ & $ 0.68$ & $ 1.26$ & $ 1.72$ \nl
$12\,36\,54.742$ & +$62\,13\,28.01$ & $16.71$ & 0.000 &
$ 2.43$ & $ 1.49$ & $ 1.19$ & $ 0.65$ & $ 0.89$ & $ 0.58$ \nl
$12\,36\,55.460$ & +$62\,13\,11.63$ & $17.07$ & 0.968 &
$73.64$ & $ 2.26$ & $ 2.08$ & $ 1.26$ & $ 1.37$ & $ 1.98$ \nl
$12\,36\,56.368$ & +$62\,12\,41.14$ & $17.72$ & 0.000 &
$ 0.76$ & $ 0.49$ & $ 0.70$ & $ 0.48$ & $ 0.61$ & $-0.04$ \nl
$12\,36\,56.675$ & +$62\,12\,45.51$ & $16.19$ & 0.518 &
$ 1.57$ & $ 2.04$ & $ 1.60$ & $ 0.83$ & $ 1.29$ & $ 1.63$ \nl
\enddata
\tablenotetext{a}{Redshifts are from Cohen et al (1996), Lowenthal et
al (1997), and Cohen et al (2000).  Sources with ``0.000'' in this
column are Galactic stars.}
\tablenotetext{b}{Colors were measured through $2.0~{\rm arcsec}$
diameter apertures in images smoothed to match the seeing of the
$3.2~{\rm \mu m}$ images; see text.  The bandpass names
$(U,B,V,I,J,H)$ stand for (F300W,F450W,F606W,F814W,F110W,F160W).
Uncertainties in all colors except $(U-B)$ and $(H-[3.2])$ are at or
below the $0.01~{\rm mag}$ level; see text}
\tablenotetext{c}{Uncertainties in $(H-[3.2])$ color are dominated
by the uncertainty in the $[3.2]_{\rm ap}$ magnitude.}
\end{deluxetable}

\begin{deluxetable}{ccccccccccccc}
\scriptsize
\tablewidth{0pt}
\tablecaption{
        Counterparts to the extragalactic $3.2~{\rm \mu m}$ sources
	\label{tab:counterparts}
}
\tablehead{
   \colhead{RA}
 & \colhead{Dec}
 & \colhead{$z$}
 & \colhead{$2.15~{\rm \mu m}$\tablenotemark{a}}
 & \colhead{$3.2~{\rm \mu m}$\tablenotemark{b}}
 & \colhead{$6.7~{\rm \mu m}$\tablenotemark{c}}
 & \colhead{$15~{\rm \mu m}$\tablenotemark{c}}
 & \colhead{$850~{\rm \mu m}$\tablenotemark{d}}
 & \colhead{$8.5~{\rm GHz}$\tablenotemark{e}}
 & \colhead{$1.4~{\rm GHz}$\tablenotemark{e}}
\\
   \multicolumn{2}{c}{(J2000)}
 & \colhead{}
 & \colhead{($\rm \mu Jy$)}
 & \colhead{($\rm \mu Jy$)}
 & \colhead{($\rm \mu Jy$)}
 & \colhead{($\rm \mu Jy$)}
 & \colhead{($\rm \mu Jy$)}
 & \colhead{($\rm \mu Jy$)}
 & \colhead{($\rm \mu Jy$)}
}
\startdata
$12\,36\,44.017$ & +$62\,12\,50.11$ & 0.557 & $ 55.9\pm 3.1$ & $ 69.2\pm 9.6$ &
         $<50$             & $282^{+60}_{-64}$ & \nodata &
         $10.2\pm 1.8$ & $<23.0$ \nl
$12\,36\,44.626$ & +$62\,13\,04.29$ & 0.485 & $ 46.5\pm 2.3$ & $ 57.6\pm 7.4$ &
         $<30$             &  $33^{+11}_{-15}$ & $3000\pm 600$~? &
         \nodata       & \nodata \nl
$12\,36\,48.088$ & +$62\,13\,09.21$ & 0.476 & $ 74.3\pm 2.7$ & $ 82.4\pm 7.6$ &
                   \nodata &           \nodata & \nodata &
         \nodata       & \nodata \nl
$12\,36\,49.435$ & +$62\,13\,46.92$ & 0.089 & $362.5\pm 5.7$ & $220.8\pm 8.1$ &
          $41^{+66}_{-30}$ & $<52$             & \nodata &
         \nodata       & \nodata \nl
$12\,36\,49.541$ & +$62\,14\,06.85$ & 0.752 & $ 21.8\pm 1.5$ & $ 35.3\pm 5.9$ &
         $<40$             & $150^{+74}_{-48}$ & \nodata &
         \nodata       & \nodata \nl
$12\,36\,51.783$ & +$62\,13\,53.85$ & 0.557 & $ 48.7\pm 2.5$ & $ 53.0\pm 7.3$ &
         $<39$             & $151^{+74}_{-68}$ & \nodata &
         \nodata       & \nodata \nl
$12\,36\,53.916$ & +$62\,12\,54.26$ & 0.642 & $ 56.9\pm 2.7$ & $ 60.8\pm 6.2$ &
         $<36$             & $179^{+60}_{-43}$ & \nodata &
         \nodata       & \nodata \nl
$12\,36\,55.460$ & +$62\,13\,11.63$ & 0.968 & $ 39.4\pm 2.4$ & $ 49.2\pm 8.2$ &
         $<37$             &  $23^{+10}_{-11}$ & \nodata &
         $12.3\pm 1.8$ & $<23.0$ \nl
$12\,36\,56.675$ & +$62\,12\,45.51$ & 0.518 & $109.5\pm 4.1$ & $110.7\pm 9.2$ &
                   \nodata &           \nodata & \nodata &
         \nodata       & \nodata \nl
\enddata
\tablenotetext{a}{Flux densities at $2.15~{\rm \mu m}$ are from Hogg
et al (2000) converted to Jy under the assumption that $K_s=0~{\rm
mag}$ corresponds to $710~{\rm Jy}$.}
\tablenotetext{b}{Flux densities at $3.2~{\rm \mu m}$ are $[3.2]_{\rm
total}$ magnitudes converted under the assumption that $[3.2]=0~{\rm
mag}$ corresponds to $330~{\rm Jy}$.}
\tablenotetext{c}{Flux densities at $6.7$ and $15~{\rm \mu m}$ are
from Aussel et al (1999).}
\tablenotetext{d}{Flux densities at $850~{\rm \mu m}$ are from Hughes
et al (1998).  The value is marked with a question mark because there
is more than one possible 1.4~GHz counterpart.}
\tablenotetext{e}{Flux densities at $8.5$ and $1.4~{\rm GHz}$ are from
Richards et al (1998) and Richards (1999b).}
\end{deluxetable}

\clearpage

\clearpage

\clearpage

\end{document}